\documentclass[5p]{elsarticle}

\usepackage{graphicx}
\usepackage{amssymb}
\usepackage{color}
\usepackage{epsfig}

\biboptions{sort&compress}

\begin{document}
\begin{frontmatter}

\title{Hunting up low-mass bosons from the Sun using HPGe detector}

\author{R. Horvat}
\author{D. Kekez\corref{cor}}
\cortext[cor]{Corresponding author.}
\ead{Dalibor.Kekez@irb.hr}
\author{M. Kr\v{c}mar}
\author{Z. Kre\v{c}ak}
\author{A. Ljubi\v{c}i\'{c}}

\address{Rudjer Bo\v{s}kovi\'{c} Institute, P.O.Box 180, 10002 Zagreb,
             Croatia}

\begin{abstract}
In this experiment we aim to look for keV-mass bosons emitted from the Sun,
by looking at a process analogous to the photoelectric/Compton effect
inside the
HPGe detector. Their coupling to both electrons and nucleons is assumed.
For masses above 25 keV, the mass dependence of our 
limit on the scalar-electron coupling
reveals a constraint which proves
stronger than that obtained recently and based on the very good agreement
between the measured and predicted solar neutrino flux from the $\rm ^{8}B$
reaction. On the other hand, the mass dependence of our 
limit on the scalar-proton/electron coupling 
together entails a limit on a possible Yukawa addition to the gravitational
inverse square low. Such a constraint on the Yukawa interactions proves
much stronger than that derived from the latest AFM Casimir force measurement.
\end{abstract}


\end{frontmatter}


The existence of new light particles (notably in the sub-eV regime), coupled
extremely weakly to ordinary
matter, seems quite feasible in extensions of the Standard Model to very
high energy scales. Besides exploring the TeV scale, currently underway via
the Large Hadron Collider at CERN, there are also overwhelming reasons (like
neutrino masses and
the dark energy of the universe) to
search for new physics in low energy experiments. Amongst the light bosonic
4-dimensional particles
inherently connected to physics at very high energy scales, the QCD axion with
pseudoscalar coupling \cite{Peccei77}, axion-like particles (like string moduli)
\cite{Conlon06} as well as  gauge bosons from the hidden sector of
string theory \cite{Aldazabal00}, leap out. Examples of bosons with scalar coupling
are the scalar components of the flavon fields \cite{Feruglio09}, the scalar
familion \cite{Reiss82} and the
sgoldstino \cite{Gorbunov01}.

Yet another motivation to study hypothetical bosons (this time with keV
masses) is the possibility that such particles may account for a dark matter
in the universe \cite{Pospelov08}. Moreover, they may account for a dark matter
component in the galactic halo as well, arising thus
naturally as a solution to the observed
model-independent annual modulation signal in the DAMA/Libra experiment
\cite{Bernabei08}\footnote{Although the interpretation of the DAMA annual modulation
in terms of keV-mass pseudoscalars is no longer viable, the solution in the
scalar case coupled to nucleons still does apply \cite{Bernabei06}.}. Besides, the
theoretical paradigm for the formation of structure in
the universe needs a keV-mass particle for the galactic scale \cite{Berezinsky93}.

Finally, a nonzero bosonic coupling to ordinary matter would induce a
violation of the Newtonian Inverse Squared Law (ISL) through new
boson-exchange forces. Such contribution gives forces between unpolarized
bodies
that violates the Weak Equivalence Principle and in
experimental tests of the gravitational ISL can be distinguished from forces
arising in higher-dimensional theories \cite{Hoyle04}.

In the present paper, we aim to observe light boson particles with mass in
the keV range, coming from the Sun and emitted in a Compton-like process
$\gamma + {\rm f} \rightarrow {\rm f} +\phi$ with the solar-plasma constituents,
where ``f'' designates electrons (e) or protons (p)\footnote{The contribution
of the Compton-like process with helium nuclei is also taken into account
when the solar flux due to a scalar-proton coupling is calculated.}. Although a
direct production via bremsstrahlung process ${\rm e} + {\rm f} \rightarrow {\rm e} +
{\rm f} + \phi$ turns out to give three times larger total emission in the Sun
\cite{Gondolo09}, the Compton-like process has a much harder spectrum for energies
$\gtrsim \mathcal{O({\rm keV})}$. Thus, for keV mass particles, the use of the
Compton-like process alone will suffice. Our experimental setup has been
designed such as to capture scalars from the Sun in the HPGe detector, in
the photoelectric-like process (if scalars couple only to electrons) or in
the Compton-like process (if scalars couple only to protons). Then we compare
our limit on the interaction strength $g_{\phi{\rm ee}}(m_{\phi})$ with that
obtained recently from the SNO constraint on nonstandard energy losses
\cite{Gondolo09}.
Also, our limit $g_{\phi{\rm pp}}(m_{\phi})$ is used to infer constraints on the
Yukawa interactions in the parameter range covered by Casimir/van der Waals
force measurements \cite{Fischbach01}.

The scalar bosons couple to ordinary matter via
formula 
\begin{equation} \label{lagrangian}
   \mathcal{L}=g_{\phi{\rm ff}}\,\phi\,\bar{\psi_{\rm f}}\,\psi_{\rm f}\,.                               
\end{equation}
In the limit of nonrelativistic electrons (protons) with mass $m_{\rm f} \gg E_{\gamma}$ 
(photon energy), the
cross section for the Compton-like process is expressed as~\cite{Gondolo09}
\begin{equation} \label{cross_section}
\sigma_{\gamma{\rm f}\rightarrow{\rm f}\phi}=\frac{1}{2}
\frac{g_{\phi{\rm ff}}^2}{4\pi\,\alpha}\,\sigma_{\rm Th}^{\rm f}\times \beta^3,
\end{equation}
where $\sigma_{\rm Th}^{\rm f}$ is the cross section for Thomson scattering 
($\sigma_{\rm Th}^{\rm e}=6.65\times 10^{-25}$~cm$^2$ and 
$\sigma_{\rm Th}^{\rm p}=1.97\times 10^{-31}$~cm$^2$), $\beta=\sqrt{1-(m_{\phi}/E)^2}$ is 
the velocity of the outgoing $\phi$-boson,
and $\alpha \simeq 1/137$ is the fine structure constant (natural units with $\hbar=c=k_{\rm B}=1$
are used through this paper).
An appreciable amount of the Sun's protons are bound inside
$^4$He nuclei and these protons have to be taken into account.
Eq.~(\ref{cross_section}), being nonrelativistic, is valid for
the $\gamma$-$^4$He scattering too
(with $\sigma_{\mbox{\rm\scriptsize Th}}^{\mbox{\rm\scriptsize He}}
=5\times 10^{-32}$~cm$^2$ and
$g_{\phi{\rm ff}}\equiv g_{\phi{\rm pp}} $).
The recoil effects are neglected so that the photon and 
the scalar energies are taken
to be equal, $E_{\gamma}=E$, where $E$ is the total energy of the scalar particle.
The differential solar flux of scalar bosons at the Earth is given by
\begin{equation}  \label{Compton}
   \frac{d\Phi_{\phi}}{dE}= \frac{1}{4\pi d_{\odot}^2}
                         \int_0^{R_\odot}dr\, 4\pi r^2 \,
        n_{\gamma}(E_{\gamma})\,\sigma_{\gamma{\rm f}\rightarrow{\rm f}\phi}\,N_{\rm f}\:.
\end{equation}
Here
\begin{equation}
n_\gamma(E_\gamma) =
\frac{E_\gamma^2}{\pi^2(e^{E_{\gamma}/T}-1)}
\end{equation}
is the differential number density of photons in the Sun
(number of photons per energy interval per volume), 
$d_{\odot}$ is the average Sun-Earth distance, while $R_\odot$ is the solar radius.
The electron, proton, and $^4$He number densities
$N_{\mbox{\rm\scriptsize e}}$, $N_{\mbox{\rm\scriptsize p}}$,
and $N_{\mbox{\rm\scriptsize He}}$, respectively,
as well as the temperature $T$ are functions of $r$,
the distance from the solar center.
They are calculated using
BS05 Standard Solar Model data \cite{Bach05}.
%
   \begin{figure}[ht]
\centerline{\includegraphics[width=80mm,angle=0]{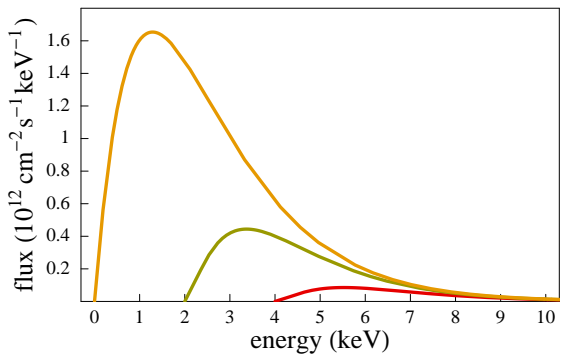}}
               \caption{Flux of solar scalars at the Earth, differential with respect to
the total energy, derived by
integrating Eq.~(\ref{Compton}) over SSM~\cite{Bach05} for
scalar masses of 0~keV (ocher line), 2~keV (green line), and 4~keV (red line).
A scalar-electron coupling of $10^{-13}$ is assumed.}
      \label{fluxe}
   \end{figure}
%
For example, Fig.~\ref{fluxe} shows differential flux of the solar scalars
at the Earth for different scalar masses and the scalar-electron 
coupling strength of 10$^{-13}$.
When we consider the flux due to the scalar-proton interactions,
we consider a sum of the proton and $^4$He contribution in
Eq.~(\ref{Compton}).

In this paper, we report results of our search for scalars which could be
produced in the Sun by the Compton-like processes and detected in the single
spectrum of an HPGe detector as a result of the photoelectric-like effect (if 
scalars couple only to electrons) or the Compton-like effect (if scalars
couple only to protons) on germanium atoms. The X-rays originating from
the Compton-like process with protons or accompanying the photoelectric-like effect
will be (subsequently) absorbed in the same crystal, and the energy of the
particular outgoing signal equals the total energy of the incoming scalar.
The expected number of events in the detector, differential with regard to the 
scalar energy $E$ is given as
\begin{equation}  \label{signal}
   \frac{dN_{\phi}}{dE}= \frac{d\Phi_{\phi}}{dE}\,
    \sigma_{\phi\,{\rm Ge}\rightarrow{\rm Ge}^{'}{\rm e}\,({\rm Ge}\,\gamma)}\,N_{\rm Ge}\,t\:,
\end{equation}
where $N_{\rm Ge}$ is the number of germanium atoms in the detector and $t$ is the
data collection time. The cross section for the photoelectric-like effect in Eq.~(\ref{signal})
is calculated from the photoabsorption cross section $\sigma_{\gamma\,{\rm Ge}\rightarrow
{\rm Ge}^{'}{\rm e}}$ as
\begin{equation}  \label{electric}
    \sigma_{\phi\,{\rm Ge}\rightarrow{\rm Ge}^{'}{\rm e}}=\frac{g_{\phi{\rm ee}}^2}
     {4\pi\,\alpha}\,\sigma_{\gamma\,{\rm Ge}\rightarrow{\rm Ge}^{'}{\rm e}}
     \times \beta^2\:,
\end{equation}
with data for $\sigma_{\gamma\,{\rm Ge}\rightarrow{\rm Ge}^{'}{\rm e}}$ 
taken from Ref.~\cite{Berger05}, while the cross section for the Compton-like process
is given by
\begin{equation}  \label{proton}
    \sigma_{\phi\,{\rm Ge}\rightarrow{\rm Ge}\,\gamma}=\frac{g_{\phi{\rm pp}}^2}
     {4\pi\,\alpha}\,\sigma_{\rm Th}^{\rm Ge}\times \beta\:,
\end{equation}
where $\sigma_{\rm Th}^{\rm Ge}=\sigma_{\rm Th}^{\rm e}(m_{\rm e}/m_{\rm Ge})^2
Z_{\rm Ge}^2$ is the Thomson cross section for a Ge nucleus.

\begin{figure}[ht]
\centerline{\includegraphics[width=80mm,angle=0]{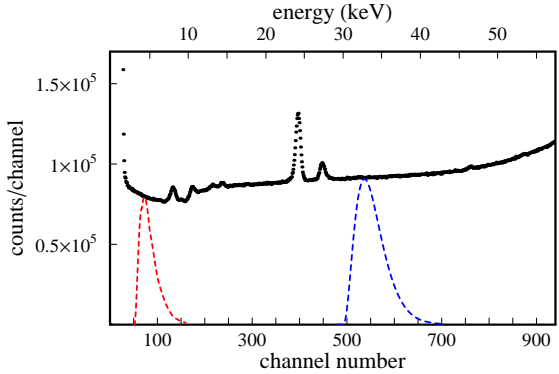}}
               \caption{Total measured energy spectrum showing also X-ray
peaks from various materials. The red and blue dashed line correspond 
to the maximum of expected events
due to scalar-electron interactions for scalar masses of 3 and 30~keV,
respectively.}
      \label{spectrum}
   \end{figure}

Because in our experimental set-up the target and the detector are the same, the 
efficiency of the system for the expected signal is $\approx$1. 
The HPGe detector with an active target mass of 1.5~kg   
was placed at ground level, inside a low-radioactivity iron box with a wall thickness 
ranging from 16 to 23~cm. The box was lined outside with 1~cm thick lead. Energy calibration
was obtained with calibrated radioactive sources of $^{241}$Am, $^{109}$Cd, and $^{55}$Fe.
For the energy region of interest in this experiment (below 60~keV) the detector resolution
was about 660~eV for the 3.9~keV escape peak and 820~eV for 13.9~keV and 59.5~keV 
gamma-rays of $^{241}$Am. Data were accumulated in a 
1024-channel analyzer, with an energy dispersion of 63.4~eV/channel
and with data collection time of $2.38\times 10^7$~s.
In these long-term running conditions, the knowledge of the energy scale is allocated by continuously
monitoring the positions and resolution of indium 24.14~keV (K$_\alpha$) and 27.26~keV 
(K$_\beta$) peaks which are present in the measured energy distribution, 	
as is seen in Fig.~\ref{spectrum}.
%
%
Drifts were $<\pm 1$ channel and a statistical accuracy of better than 0.4\% per 
channel was attained.

\begin{figure}[ht]
\centerline{\includegraphics[width=80mm,angle=0]{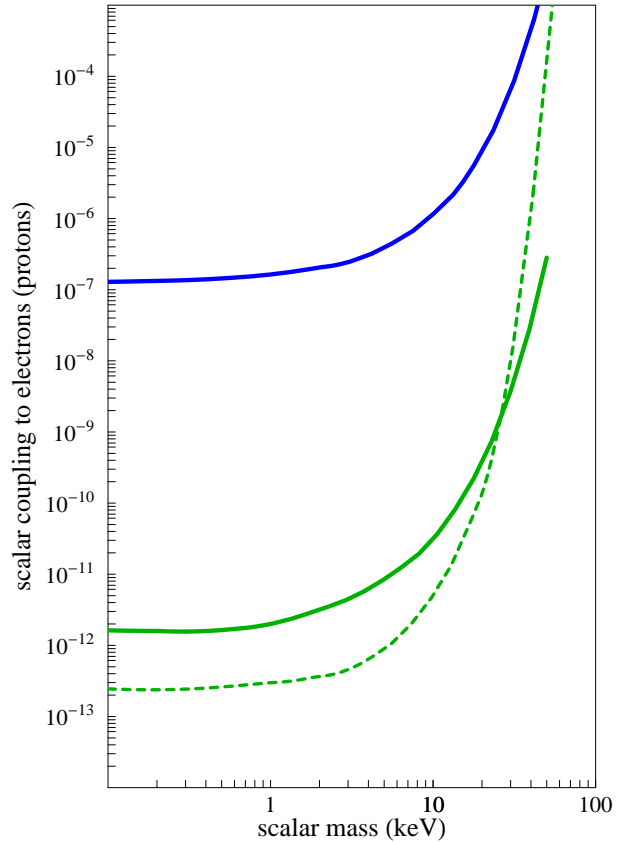}}
               \caption{Upper limits on the coupling of scalar to electrons 
(green solid line) and to protons (blue solid line), achieved by our experiment,
are shown as a function of the scalar mass. The green dashed line corresponds
to the bound on the scalar-electron coupling derived from nonstandard
solar energy losses~\cite{Gondolo09}.}
      \label{couplings}
   \end{figure}

The evaluation for upper limits on the scalar-electron (proton) coupling follows the
most conservative assumption, by requiring the predicted signal in an energy bin to 
be less than or equal to the recorded   
counts. Similar approaches have been used by other groups \cite{Bau98} in a
case like this where direct background measurement is not possible (the Sun cannot be switched
off) and the signal shape is a broad smooth spectrum on top of an unknown background spectrum.
For fixed $m_{\phi}$, the theoretically expected spectrum of scalar-induced events has been
calculated by means of Eq.~(\ref{signal}), where $g_{\phi{\rm ff}}^4$ is the only free parameter
which is then used to fit the maximal value of the expected spectrum to the measured one. 
Figure~\ref{spectrum} shows a comparison between the experimental data and the calculated
spectrum for scalar-electron interactions, for scalar masses of 3~keV (red 
dashed line) and 30~keV (blue dashed line). The corresponding upper limits on the scalar-electron
(proton) coupling strength obtained in this work 
are displayed in Fig.~\ref{couplings}. 
%
%
One can see that for masses above 25 keV, the mass dependence of our $g_{\phi {\rm ee}}$ limit
reveals a constraint which proves
stronger than that obtained recently~\cite{Gondolo09} and based on the very good agreement
between the measured and predicted solar neutrino flux from the $\rm ^{8}B$
reaction. The possibility of decay of the scalar particle into two photons, with a lifetime
of $64\pi/(g_{\phi\gamma\gamma}^2\,m_{\phi}^3)$ and $|g_{\phi\gamma\gamma}|=2\,g_{\phi{\rm ff}}\,
Z_{\rm f}^2\,\alpha/(3\pi\,m_{\rm f})$~\cite{Bernabei06}, during its traversal from
the Sun to the Earth sets an upper bound on the scalar mass to be less than about 50~keV.      
%
   \begin{figure}[ht]
\centerline{\includegraphics[width=80mm,angle=0]{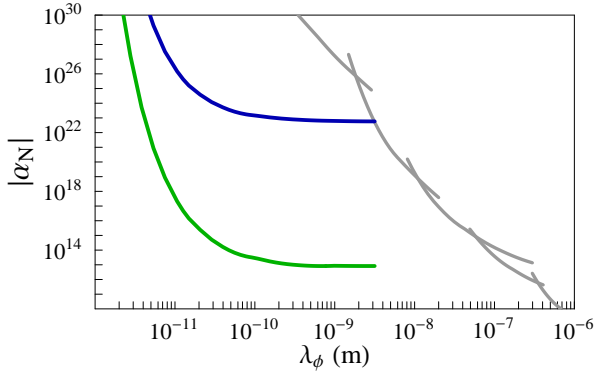}}
               \caption{Upper limits on new forces from potentials of the form
given by Eq.~(\ref{potential}).
Our results corresponding to scalar-proton and scalar-electron interactions are shown by
the blue and the green line, respectively.
The gray curves represent the limits from various van der Waals/Casimir force
measurements~\cite{Bordag94,Mostepanenko01,Fischbach01,Mostepanenko88,Bordag98}. }
\label{alpha}
   \end{figure}
%

We note that from our limit
on the extension of the Standard Model $g_{\phi{\rm ff}}(m_{\phi})$, one can
infer also a constraint on a deviation from Newtonian physics. Indeed, a
spin-0 exchange contribution to the static gravitational energy, described by
the lagrangian density (\ref{lagrangian}), is given by a Yukawa addition to the familiar
Newton potential between two point masses $m_1$ and $m_2$,
\begin{equation}
V(r) = -G \,\frac{m_1 \,m_2}{r} \left [1 + \alpha_{\rm N}\,e^{-r/\lambda_{\phi}}
\right ] \;.
\label{potential}
\end{equation}
Here $\alpha_{\rm N}$ is proportional to the squared product of the
appropriate coupling constants,
$\alpha_{\rm N} \simeq g_{\phi{\rm ff}}^2 /(16 \pi G m_{{\rm p}}^2)$. This constant
characterizes the strength of non-Newtonian interaction compared
to the Newtonian one with a gravitational constant $G=1/M_{\rm Pl}^2$. The interaction
range $\lambda _{\phi}=1/m_{\phi}$ has the meaning of the Compton wavelength of the spinless particle. The fact that matter-scalar coupling
constants in (\ref{potential}) are species-dependent, reflects a violation of the
Weak Equivalence Principle. The region in the plane $(\lambda_{\phi},|\alpha_{\rm N}|)$, 
excluded by various torsion-balance experiments as well as
theoretical prediction from string-like theories are collected in
\cite{Hoyle04}. On the other hand, since experiments searching for
$\alpha_{\rm N}$-dependent term in (\ref{potential}) are only sensitive when 
$r \sim \lambda_{\phi}$, the
experiments that are setting the best limits for separations $\lesssim 10^{-4}$~m are
those testing the Casimir force law. This is so since in that regime Casimir
forces, and not gravity, provide the dominant background force. Our limit
$g_{\phi{\rm pp}}(m_{\phi})$, when expressed in the $(\lambda_{\phi},|\alpha_{\rm N}|)$ 
plane, just fits the regime covered by Casimir force measurements.
As seen from Fig.~\ref{alpha}, the constraints
from various Casimir force measurements are
considerably weaker than ours for $\lambda_{\phi} \lesssim 10^{-8}$~m. Thus, all the
existing constraints in the plane $(\lambda_{\phi},|\alpha_{\rm N}|)$ 
for $\lambda_{\phi} \lesssim 10^{-8}$~m should be superseded by the currently
derived ones.

In conclusion, we have performed an experimental search for low-mass bosons
coming  from the Sun. For bosons coupling to protons, our limit
comfortably extends to masses beyond the solar central
temperature, encompassing thus marginally
the allowed region for the presence of scalar dark matter in the galactic
halo. Also, our limits are complementary to those obtained recently from the
accurate knowledge of the solar interior. Finally, our limits tighten
up constraints on non-Newtonian gravity much more efficiently than
those obtained from the Casimir force measurements.

The authors acknowledge the support of the Croatian MSES Project 
No. 098-0982887-2872.

\end{document}